\documentclass[10pt,letterpaper]{article}
\usepackage{opex3}

\begin{document}

\title{Temperature measurement and stabilization in a birefringent whispering gallery resonator}

\author{D.V. Strekalov, R.J. Thompson, L.M. Baumgartel, I.S. Grudinin, and N. Yu}

\address{Jet Propulsion Laboratory, California Institute of
Technology, 4800 Oak Grove Drive, Pasadena, California 91109-8099.
}

\email{Dmitry.V.Strekalov@jpl.nasa.gov}

\begin{abstract}
Temperature measurement with nano-Kelvin resolution is demonstrated at room temperature, based on the thermal dependence of an optical crystal anisotropy in a high quality whispering gallery resonator. As the resonator's TE and TM modes frequencies have different temperature coefficients, their differential shift provides a sensitive measurement of the temperature variation, which is used for active stabilization of the temperature. 
\end{abstract}

\section{Introduction}

Whispering Gallery Mode (WGM) resonators hold a unique place among optical resonators for several reasons. Relying on continuous total internal reflections for light confinement, such resonators feature extremely high quality factors $Q$ ($Q>10^{11}$ has been demonstrated \cite{Savchenkov07Q}) in a very broad optical range. They are also very compact, with a typical size ranging from a few tens of microns to several millimeters, and are entirely solid-state. It is therefore not surprising that among many other applications, WGM resonators have been suggested for use as reference cavities for laser stabilization \cite{Vassiliev98,Carmon05,Sprenger10,Liang10,Alnis11XXX}. Such an approach to laser stabilization is predicted \cite{Savchenkov07JOSAB-II} to allow for the fractional optical frequency stability at the level of one part per $10^{-14}$ at 1 s. In addition to optical clocks, laser stabilization with a WGM resonator holds a great promise for high-resolution spectroscopy, optical metrology and other fields. 

The main practical difficulty associated with using WGM resonators for laser stabilization arises from the fundamental fact that the light propagates inside an optical material. This appears to be a disadvantage compared with e.g. low-expansion Fabri-Perot resonators that are essentially vacuum-filled. In WGM resonators the material's optical properties set the limit not only for the resonator's $Q$-factor, but also for its stability. Indeed, it is easy to see that even a very small variation $\Delta (nL)$ of the resonator's optical length $nL$ such that $\Delta (nL)/(nL) = 1/Q$ causes the resonance frequency to shift by one linewidth.

The most important factors contributing to variation of $\Delta (nL)$ are the thermal refractivity and thermal expansion \cite{Savchenkov07JOSAB-II}. One approach proposed \cite{Savchenkov07JOSAB-II,Matsko11T} for compensating these variations is to use the temperature dependence of the resonator's own anisotropy. In this work we report the first practical implementation of this approach and demonstrate active temperature stabilization of a WGM resonator at 76.04 $^\circ$C. We achieve temperature stabilization at the level of 200 nK when integrated for 1 second, and below 10 nK when integrated for 10,000 seconds. We see this result as significant not only in the context of laser stabilization but also as a demonstration of a novel temperature sensor with a broad spectrum of potential applications.


\section {Theoretical estimates}

Temperature dependence of a WGM frequency $f$ can be found by differentiating the WGM dispersion equation (found in e.g. \cite{Gorodetsky06}) with respect to temperature $T$:
\begin{equation}
\frac{1}{f}\frac{df}{dT}+\frac{1}{n}\frac{dn}{dT}+\frac{1}{R}\frac{dR}{dT}=0.
\label{var}
\end{equation}
In Eq.~(\ref{var}) $R$ is the resonator radius and $n$ is the refractive index for the given polarization and wavelength $\lambda$. This equation is an approximation, where the small terms of the order $\lambda/R$ and higher have been neglected.

The second and third terms of Eq.~(\ref{var}) are equal to the thermorefractivity and thermal expansion coefficients, $\alpha_n^{(o,e)}$ and $\alpha_l^{(o)}$ respectively. Here the superscripts indicate ordinarily and extraordinarily polarized light. Since our resonator is shaped as a disk made from a z-cut magnesium fluoride (MgF$_2$) wafer, its TE and TM modes support the ordinary (horizontally) and extraordinary (vertically) polarized light, respectively. Therefore the thermorefractivity of the former is approximately equal to $\alpha_n^{(o)}$, and of the latter, to $\alpha_n^{(e)}$. The thermal expansion coefficient, on the other hand, is $\alpha_l^{(o)}$ for both mode families. These coefficients for MgF$_2$ are given by \cite{Savchenkov07JOSAB-II}:
\begin{eqnarray}
\alpha_n^{(o)}&=&(7.14-0.04062\,T)\times 10^{-7}\,K^{-1},\nonumber\\
\alpha_n^{(e)}&=&(3.02-0.04071\,T)\times 10^{-7}\,K^{-1},\nonumber\\
\alpha_l^{(o)}&=&9\times10^{-6}\,K^{-1},
\label{alphas}
\end{eqnarray}
where $T$ is in $^\circ$C. From (\ref{alphas}) we see that while the thermal variation of each mode's frequency is dominated by the thermal expansion, 
\begin{equation}
\frac{d}{dT}f_{o,e}\approx-\frac{c}{\lambda}\,\alpha_l^{(o)}\approx -1.73\;{\rm GHz/K},
\label{dL}
\end{equation}
the variation of the frequency \emph{difference} $\Delta f = f_o-f_e$ depends only on the differential thermorefractivity:
\begin{equation}
\frac{d}{dT}\Delta f=-\frac{c}{\lambda}\left(\alpha_n^{(o)}-\alpha_n^{(e)}\right)\approx -79.1\;{\rm MHz/K}.
\label{dvar}
\end{equation}
We use this temperature dependence to measure and stabilize the resonator temperature.

\section{Experimental demonstration}

\begin{figure}[b]
\centerline{
\input epsf
\setlength{\epsfxsize}{4in} \epsffile{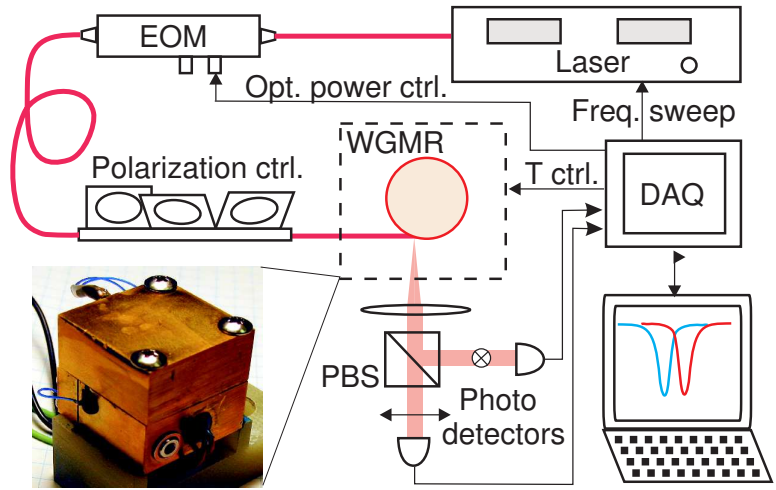} 
}
\caption[]{\label{fig:setup}The experimental setup. Temperature of the brass cube on the photo is set to 76 $^\circ$C and stabilized to a few hundred nanodegrees using the dual-mode stabilization technique.}
\end{figure}

Our experimental setup is shown in Fig.~\ref{fig:setup}. The key element of this setup is a MgF$_2$ WGM resonator which is shaped as a disk of approximately 8 mm in diameter and 2 mm thick.  Its modes are excited by a CW DFB laser at $\lambda=1560$ nm. Using a fiber optic polarization adjuster we set input light polarization such as to excite both TE and TM mode families. The optical power at the output of the polarization adjuster ranges from 0.5 to 1.5 mW in each polarization.

The light is coupled into and out of the resonator using the standard \emph{frustrated total internal reflection} technique with an angle-polished fiber \cite{Ilchenko99}. The coupling rate is controlled using a micro screw and a two-springs differential that can adjust the distance between the resonator rim and the fiber coupler to a fraction of the optical wavelength. 

The resonator and fiber coupler are mounted inside a 1 inch brass cube, shown in the photo in Fig.~\ref{fig:setup}. This cube also has two built-in resistive heaters and a thermistor to provide the initial temperature stabilization by means of a commercial temperature controller. During the active phase of our experiment this controller is disengaged and the heaters are controlled by a feedback loop described below. The cube is placed in a cardboard box on an optical table, with no other means of external temperature stabilization. 

The out-coupled light is collimated by a lens and split by a polarizing beam splitter. Two photodetectors at two ports of the beam splitter can therefore observe the TE and TM modes. The photodetectors' signals are acquired by a computer-controlled DAQ which also controls the laser frequency and performs the temperature stabilization. 

\begin{figure}[t]
\vspace*{-0.2in}
\centerline{
\input epsf
\setlength{\epsfxsize}{3.1in} \epsffile{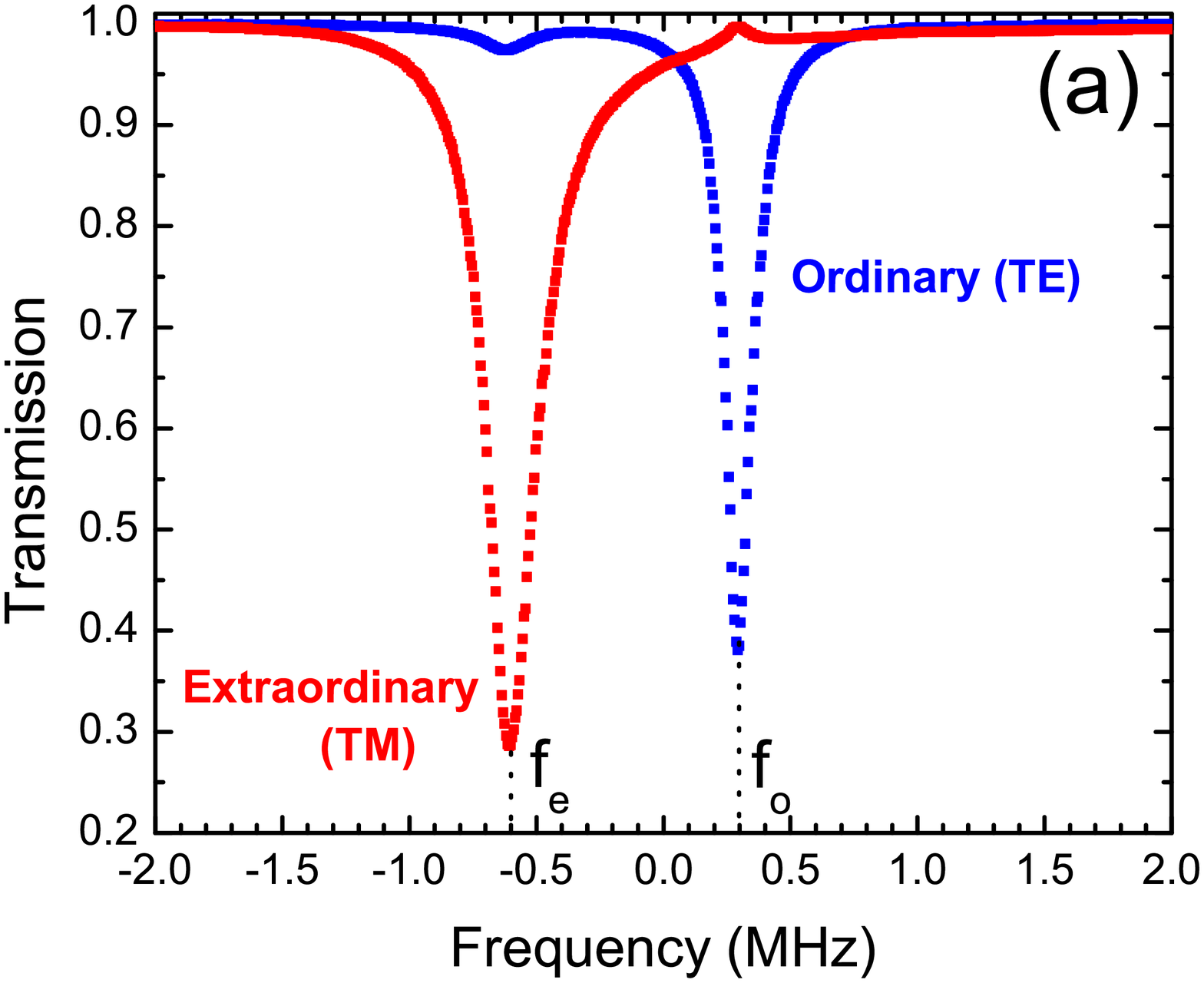} \hspace*{-0.5in}
\input epsf
\setlength{\epsfxsize}{3.1in} \epsffile{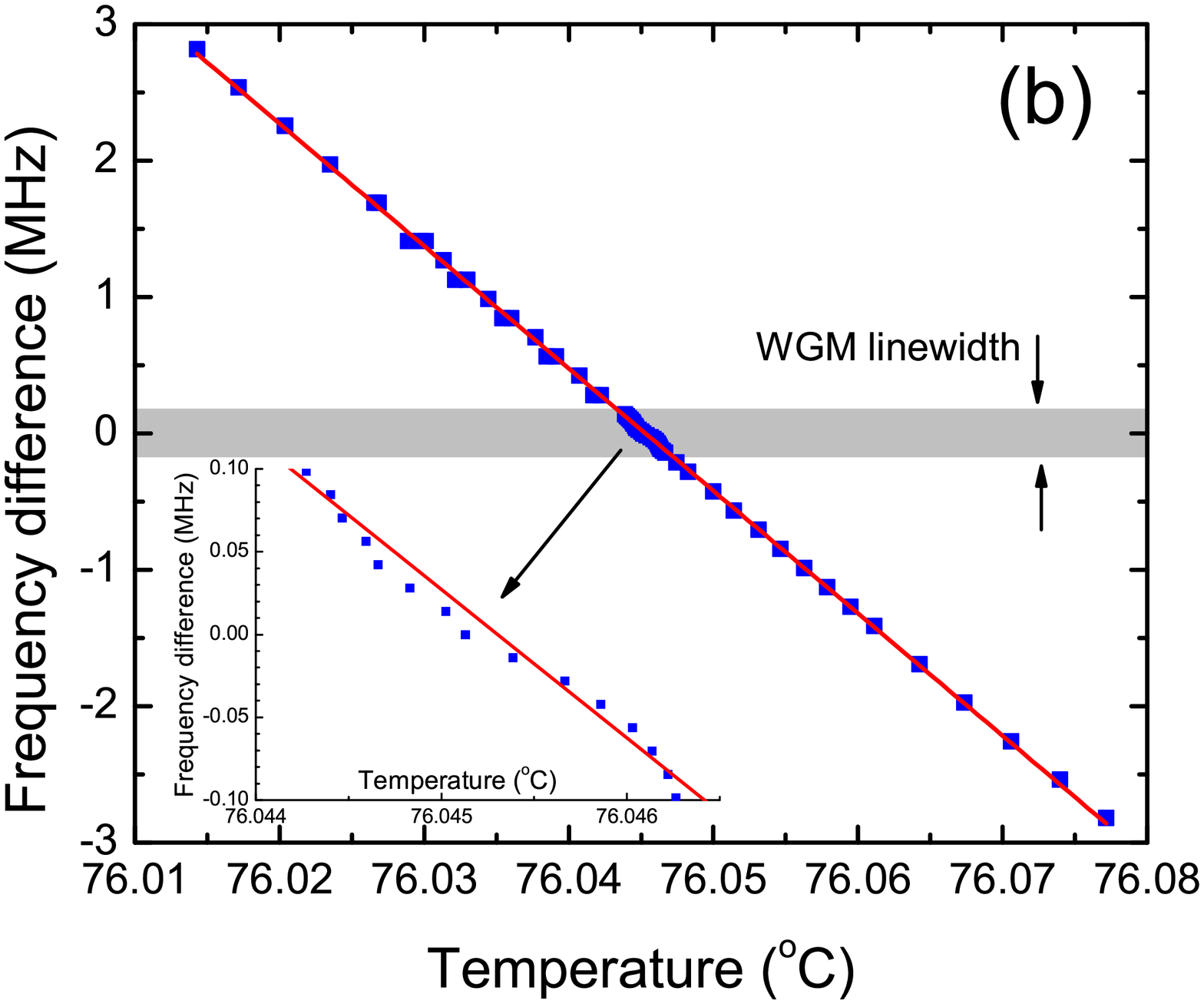} 
}
\caption[]{\label{fig:lines} (a) A pair of TE and TM WGMs with a small frequency detuning $\Delta f = f_o-f_e$.\\ (b) Temperature dependence of $\Delta f(T)$ is linear with a -89.8 MHz/K slope. The inset shows a small distortion around the set point $\Delta f(T_{set})=0$ which is due to the WGMs cross-coupling. }
\end{figure}

By varying the laser wavelength or the resonator temperature it is possible to find multiple pairs of TE and TM modes that have sufficiently close frequencies $f_o$ and $f_e$, as shown in Fig.~\ref{fig:lines}(a). In our resonator a typical intrinsic (weakly coupled) value of $Q$ for TE WGM is $2\times 10^9$, and  for TM WGM is $1\times 10^9$.

To observe the WGMs, the DAQ sweeps the laser frequency 100 times per second, each time acquiring 2500 data points per channel. This data is analyzed in real time during each sweep, and the WGM frequencies $f_o$ and $f_e$ are found by peak fitting. Their sum is used to continuously re-center the laser frequency sweep with respect to the modes. Their difference $\Delta f = f_o-f_e$ is used as the temperature error signal. This signal has been calibrated by measuring $\Delta f$ simultaneously with the resistance of the thermistor, which required a seven-digit Ohm meter. The calibration curve shown in Fig.~\ref{fig:lines}(b) yields the temperature tuning rate of $d(\Delta f)/dT = -89.8\pm 0.2$ MHz/K, which is reasonably close to the theoretical value found in (\ref{dvar}). 

To demonstrate the use of our temperature measurement technique for temperature stabilization, we have implemented two Labview-based feedback loops with proportional and integral gains. The slower loop uses the heaters mounted inside the brass cube to provide the long-term stabilization. The faster loop uses an amplitude electro-optical modulator to vary the optical power, which has a high-speed heating effect \cite{Fomin04,Carmon04,DelHaye08}. The proportional and integral gains of both feedback loops, as well as the cross-over time of the loops, are optimized for the best short term and long term performance.

\begin{figure}[t]
\vspace*{-0.2in}
\centerline{
\input epsf
\setlength{\epsfxsize}{3.1in} \epsffile{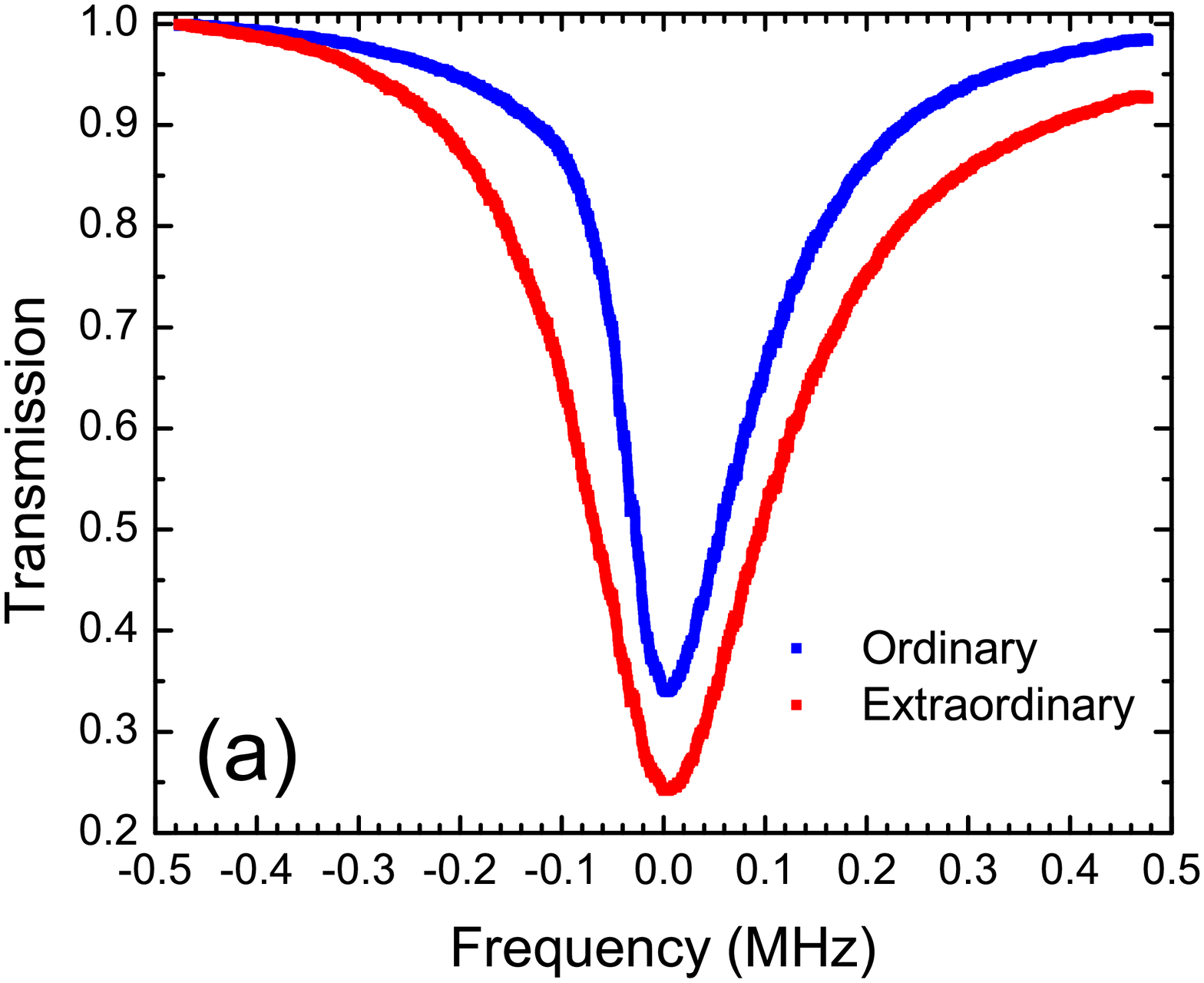} \hspace*{-0.5in}
\input epsf
\setlength{\epsfxsize}{3.1in} \epsffile{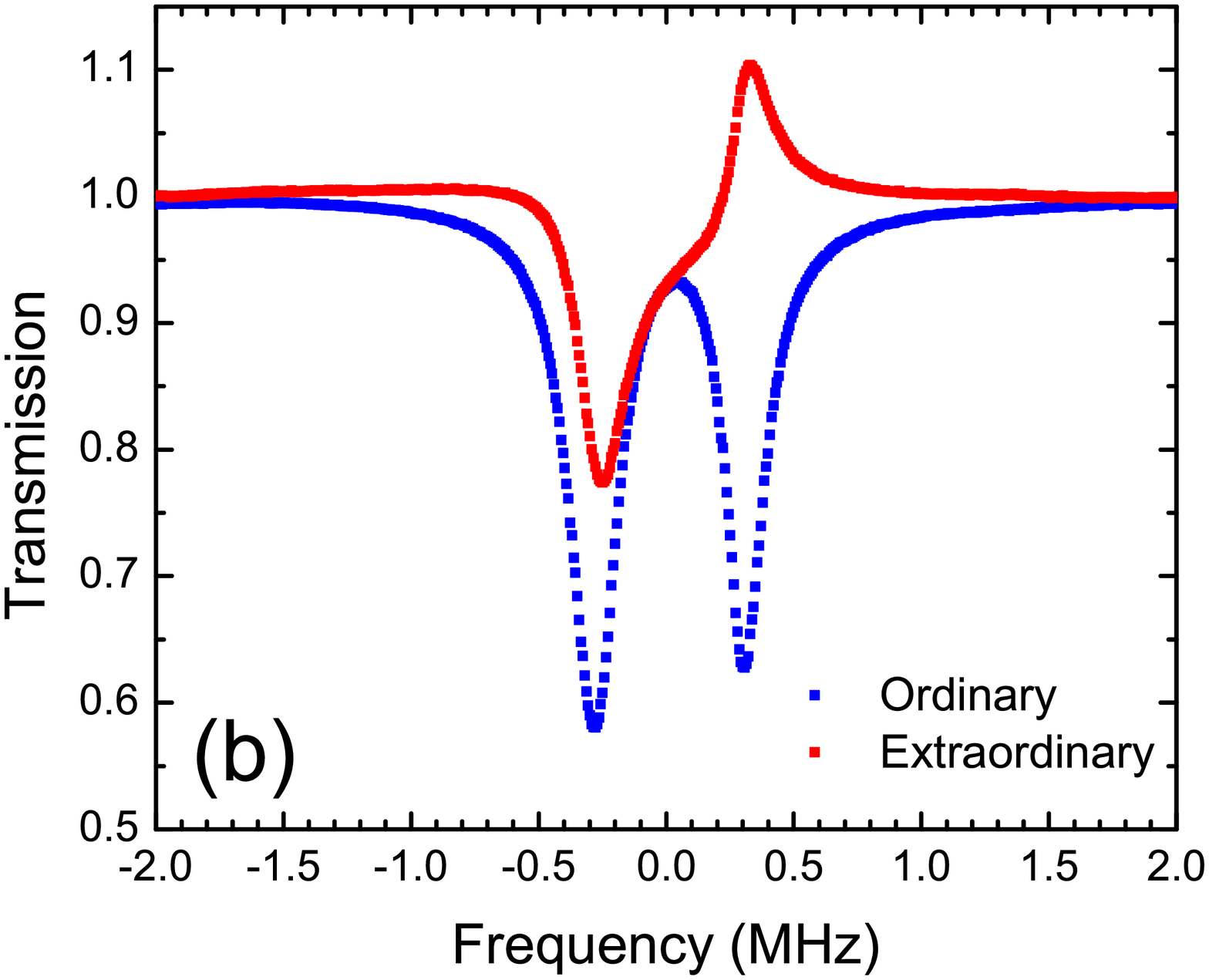} 
}
\caption[]{\label{fig:locked2} (a) The WGMs shown in Fig.~\ref{fig:lines} are brought on-resonance by temperature control. These modes have weak nonlinear coupling. (b) Another pair of modes couples strongly and is not suitable for our application.}
\end{figure}

When the temperature control loops are closed, the TE and TM modes shown in Fig.~\ref{fig:lines}(a) are brought on resonance $\Delta f(T_{set})=0$ as shown in Fig.~\ref{fig:locked2}(a), and the resonator temperature is stabilized at that point.  Special care needs to be taken to choose the WGMs that have weak cross-coupling, which may arise from either linear or nonlinear interaction between the WGMs and can severely distort the WGM spectrum (see Fig.~\ref{fig:locked2}(b)). Linear cross-coupling may be caused by inhomogeneities, such as e.g. surface scratches or dust. Such defects are not usually expected in high-$Q$ resonators. Nonlinear coupling may be due to thermal, acoustical or Kerr nonlinearities, and is directly related to the WGMs spatial overlap. 

The modes shown in Fig.~\ref{fig:locked2}(a) interact only weakly. To verify that this interaction does not significantly alter the linear dependence $\Delta f(T)$ near the set point, we took a number of data points within the WGM linewidth. These data points are shown in the inset in Fig.~\ref{fig:lines}(b). With such a fine step we should be able to capture all reasonably expected distortions. We see that while the cross-coupling effect is clearly present, it is not likely to strongly affect our measurements.

\begin{figure}[htp]
\centerline{
\input epsf
\setlength{\epsfxsize}{4.2in} \epsffile{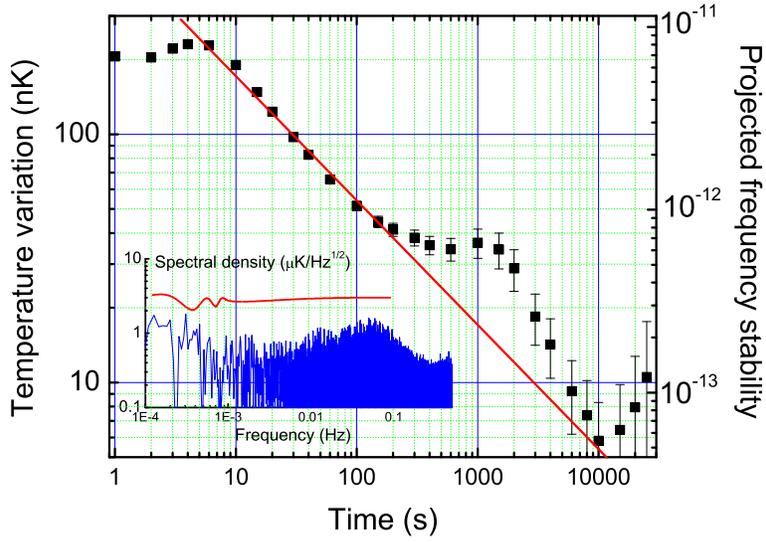} 
}
\caption[]{\label{fig:AD} The Allan variance of the temperature fluctuations in the stabilized WGM resonator in nano-Kelvin and in terms of the expected fractional frequency stability. The straight line has a slope of 0.5 $\mu$K / s$^{1/2}$. The error bars are based on the measurements statistics. On the inset: the spectral density of the temperature fluctuations measured in our system (below) and the noise-equivalent temperature spectral density in \cite{Sanjuan07} (above).}
\end{figure}

 
To characterize the performance of our temperature stabilization setup, we record a time series of the $\Delta f$ fluctuations and find both their spectral density and Allan variance. These are converted to the temperature units using the experimental value for $d(\Delta f)/dT$. A typical result is shown in Fig.~\ref{fig:AD}. The optical powers injected in the WGM resonator in this case were 0.8 mW for TE and 1.1 mW for TM polarizations. The peculiar non-monotonic behavior of the Allan variance below 5 seconds that can be seen in this Figure is a consequence of a very high integral gain set in the slow control loop, which made it nearly unstable. The high gain was necessary to provide a good long-term stability against the ambient temperature variations to which our system has been considerably exposed. The residual signature of these variations is visible around 1000 seconds and is likely to be related to the building air conditioner, which has an appropriate period of operation. Apart from this signature, the Allan variance in Fig.~\ref{fig:AD} scales as 0.5 $\mu$K / s$^{1/2}$ for the averaging time longer than 5 seconds. The inverse square-root behavior of the Allan variance has been theoretically predicted in \cite{Savchenkov07JOSAB-II}.  

For comparison, we measured the temperature Allan variance of the resonator stabilized by the commercial digital temperature controller, and found it roughly flat at a level of 0.2 to 0.7 mK. Therefore, the temperature stabilization technique we report outperforms this controller by over three orders of magnitude with only 1 second integration time. It is also interesting to compare our results on temperature stabilization with sensitivity of the state-of-art non-cryogenic temperature sensors (which are sensing but not controlling temperature) developed for LISA \cite{Sanjuan07}. This sensitivity is given as the noise-equivalent temperature spectral density, which is also shown in the inset of Fig.~\ref{fig:AD}.  
From this we see that our temperature stabilization technique considerably outperforms the LISA sensors. 

The temperature sensitivity of our approach is apparently limited by the accuracy of determining the WGM resonance frequencies. A least-square Lorentzian fitting of the peaks such as shown in Fig.~\ref{fig:locked2}(a) yields the standard error of their frequency difference which corresponds to $\sigma_{\Delta f}(10\,{\rm ms})\approx 4.8\,\mu$K. With 100 such measurement per second we may expect $\sigma_{\Delta f}(1\,{\rm s})\approx 480$ nK. The observed one-second Allan variance of less than a half of this value indicates that we are at or near the accuracy limit of our measurement. It should be pointed out that the WGM peaks in Fig.~\ref{fig:locked2}(a) are noticeably asymetric due to the dynamic heating by the laser sweep \cite{Fomin04,Carmon04,DelHaye08}. Therefore reducing the sweep time with a faster DAQ would not only increase the data rate but also improve each peak fitting accuracy by suppressing this asymetry, and will improve the temperature sensitivity of our method. 

Since the main motivation of the demonstrated temperature stabilization technique is to provide a stable laser locking reference, it is interesting to evaluate the projected fractional frequency stability of such a reference. This can be done by multiplying the measured temperature stability by the coefficient (\ref{dL}). This result is also shown in Fig.~\ref{fig:AD}. According to this, we should be able to stabilize our laser to one part per $7\times 10^{-12}$ per 1 s integration time, and down to one part per $6\times 10^{-14}$ per 10,000 s integration time even with the present proof-of-principle system. 

It should be pointed out that in addition to its temperature dependence, the dual-mode frequency detuning $\Delta f$ may have other dependencies, e.g. on the atmospheric pressure, TE/TM power ratio, etc. It also could be time-dependent due to crystal aging. Successfully using $\Delta f$ for temperature stabilization however shows that its temperature dependence is dominant, and serves as a validation of our technique of ultra-sensitive temperature measurements. A further verification of the achieved temperature stability will be available from measuring the relative frequency stability of two independently stabilized lasers, which will be the subject of our future work.

\section{Conclusions}
 
We have demonstrated a WGM based temperature sensor with nano-Kelvin sensitivity operating at room temperature. We have used it to stabilize a WGM resonator at the level of a few nano-Kelvin, which will allow us to use this resonator as an ultra-stable laser lock reference. The demonstrated technique can be used in a variety of other applications requiring high temperature stability, as well as ultra-sensitive measurements of temperature variations. As a few examples of such applications we would like to suggest the thermal stabilization of quartz oscillator, mid- or far-IR sensitive bolometers, precise calorimetric measurements in chemistry, and study of optical and mechanical aging effects in various crystalline resonators. 

\section{Acknowledgements}

The research described in this paper was carried out at the Jet
Propulsion Laboratory, California Institute of Technology, under a
contract with the NASA. 

\end{document}